# Ultra-Low Threshold Monolayer Semiconductor Nanocavity Lasers

Sanfeng Wu[1], Sonia Buckley[2], John R. Schaibley[1], Liefeng Feng[1,3], Jiaqiang Yan[4,5], David G. Mandrus[4,5,6], Fariba Hatami[7], Wang Yao[8], Jelena Vučković[2], Arka Majumdar[9,*], Xiaodong Xu[1,10,*]

[1]Department of Physics, University of Washington, Seattle, Washington 98195, USA
[2]Ginzton Laboratory, Stanford University, Stanford, CA 94305, USA
[3]Department of Applied Physics, Tianjin University, Tianjin 300072, China
[4]Materials Science and Technology Division, Oak Ridge National Laboratory, Oak Ridge, Tennessee, 37831, USA
[5]Department of Materials Science and Engineering, University of Tennessee, Knoxville, Tennessee, 37996, USA
[6]Department of Physics and Astronomy, University of Tennessee, Knoxville, Tennessee 37996, USA
[7]Department of Physics, Humboldt University, D-12489, Berlin, Germany
[8]Department of Physics and Center of Theoretical and Computational Physics, University of Hong Kong, Hong Kong, China
[9]Department of Electrical Engineering, University of Washington, Seattle, Washington 98195, USA
[10]Department of Material Science and Engineering, University of Washington, Seattle, Washington 98195, USA

**Engineering the electromagnetic environment of a nanoscale light emitter by a photonic cavity can significantly enhance its spontaneous emission rate through cavity quantum electrodynamics in the Purcell regime. This effect can greatly reduce the lasing threshold of the emitter[1–5], providing the ultimate low-threshold laser system with small footprint, low power consumption and ultrafast modulation. A state-of-the-art ultra-low threshold nanolaser has been successfully developed though embedding quantum dots into photonic crystal cavity (PhCC)[6–8]. However, several core challenges impede the practical applications of this architecture, including the random positions and compositional fluctuations of the dots[7], extreme difficulty in current injection[8], and lack of compatibility with electronic circuits[7,8]. Here, we report a new strategy to lase, where atomically thin crystalline semiconductor, i.e., a tungsten-diselenide ($WSe_2$) monolayer, is nondestructively and deterministically introduced as a gain medium at the surface of a pre-fabricated PhCC. A new type of continuous-wave nanolaser operating in the visible regime is achieved with an optical pumping threshold as low as 27 nW at 130 K, similar to the value achieved in quantum dot PhCC lasers[7]. The key to the lasing action lies in the monolayer nature of the gain medium, which confines direct-gap excitons to within 1 nm of the PhCC surface. The surface-gain geometry allows unprecedented accessibilities to multi-functionalize the gain, enabling electrically pumped operation. Our scheme is scalable and compatible with integrated photonics for on-chip optical communication technologies.**

Monolayer transitional metal dichalcogenides (TMDCs) with chemical formula $MX_2$ (M=W, Mo; X=S, Se, Te. See Fig. 1**a** for crystal structure) are the first class of two-dimensional (2D) semiconductors with a direct band gap in the visible frequency range[9,10], where tightly bound excitons with novel properties are discovered[11–13]. These structurally stable, mechanically strong, electrically tunable, and optically active materials have generated vast interest in the scientific community due to their great potential for spin-valleytronics[14,15], field effect transistors[16], light emitting diodes[17,18,19], solar cells[20] and photo-detectors[21], expanding new realms in 2D-crystal-based science and device applications.

We here demonstrate the first nanoscale laser system based upon 2D quantum materials, harnessing the unique advantages of atomically thin crystals for coherent light generation. In our architecture, monolayer $WSe_2$, as seen in the optical image in Fig. 1**b**, is selected as the gain medium due to its desirable bandwidth and relatively high photoluminescence (PL) quantum yield compared to other TMDC monolayers. The monolayer is coupled to a prefabricated PhCC on a gallium phosphide (GaP) thin membrane[22] which is transparent to $WSe_2$ emission (See Fig.1**a** and Methods). An L3 type of PhCC is employed[23], where three neighboring holes in linear arrangement are missing, shown by the scanning electron microscopy (SEM) image in Fig. 1**c**. The PhCC is carefully designed to yield the highest quality factor (Q-factor) mode with resonant energy around 740nm, which is in the band of the monolayer PL.

Controlled spontaneous emission was recently demonstrated in monolayer semiconductors, where low Q-factor PhCCs[24,25] (~300) or distributed Bragg reflectors[26] were used. In our devices, the as-fabricated PhCCs have Q-factors of ~$10^4$ (Extended Data Fig. 1), improved by ~30 times. This results in a significant improvement of the Purcell factor[24,25] (Supplementary Information S1), which is crucial for lasing. To achieve such a high Q-factor in the visible region, we use a 125 nm thick membrane (see Methods), which is 55 nm thinner than our previously reported low-Q cavity where no lasing behavior was observed[25]. This design significantly improves the cavity Q-factors, due to an optimal thickness to lattice constant ratio, and more importantly, an improved sidewall-verticality due to the lower aspect ratio of the etched holes. Conical (non-vertical) etching of the holes leads to coupling to leaky TM modes of the slab[27], which eventually decreases the Q.

The gain-cavity coupling is achieved through directly transferring WSe$_2$ monolayer on top of the PhCC, using methods that are well established for 2D materials. In the cartoon plot of Fig. 1**a**, we show the electric-field intensity profile (x-y plane) of the fundamental mode defined by our cavity, simulated by finite-difference time-domain (FDTD) method[23]. Fig. 1**d** illustrates the cross section (x-z plane) profile of the mode, where the orange dashed line indicates the ideal position of WSe$_2$ monolayer. The corresponding electric field intensity at the monolayer maintains ~ 40% of the maximum located at the center, allowing for efficient overlap between the cavity mode and monolayer WSe$_2$ on the surface. In our geometry, even though the gain medium is placed outside the cavity, the ultimate miniaturization in thickness (~0.7nm) of the monolayer allows the minimal degradation of gain-cavity coupling.

Lasing at a reduced threshold power is achieved by enhancing spontaneous emission into a resonant cavity mode. Fig. 1**e** shows a typical emission spectrum of the hybrid structure, taken under optical pumping by 632 nm continuous-wave (CW) laser at 80 K. The laser emission is the sharp feature located at 739.7 nm. We extract 0.3 nm linewidth at the half maximum of this spectrum. The peak is polarized in the "y" direction, consistent with the fundamental mode of the cavity.

One hallmark feature of a laser is the nonlinear "kink" that occurs around the lasing threshold in the log scale plot of the output light intensity (detected power by integrating over the spectrum) as a function of pump (*L-L* curve). In Fig. 2**a-b**, we present the *L-L* curves (red filled square) for the monolayer laser at temperatures of 130 K and 80 K, both showing the nonlinear "kink" at the laser threshold region. We estimate the typical emission power levels, after the objective, of our lasing devices at this region to be ~ 10 fW with 100 nW incident pump power. A set of power-dependent data for spontaneous emission off cavity resonance is also shown as contrast (violet half-filled square), where no "kink" signature is observed. The PL spectra corresponding to the denoted data points (arrows in Fig. 2**b**) are shown in Fig. 2**c**. The *L-L* curve is fitted by cavity laser rate equation (Supplementary Information S2), as shown by the solid lines.

In a nanocavity laser, the $\beta$-factor is the figure of merit that characterizes the laser threshold, and is defined as the fraction of the spontaneous emission into the cavity mode (Supplementary Information S2). A large $\beta$-factor reduces the lasing threshold power. We find that $\beta = 0.19$ is the best fit to our observed data, while $\beta = 0.05$ and $\beta = 1$ are also plotted for reference. This indicates that in our WSe$_2$-cavity system ~19% of the total spontaneous emission is coupled to the cavity mode, comparable to the performance achieved in quantum-dot photonic crystal cavity lasers. We extract the lasing threshold of our device to be 27 nW (~1 W/cm$^2$) measured by the incident power. Such ultralow threshold lasing behavior demonstrates that the cavity-gain coupling in the surface-gain geometry is as efficient as the embedded quantum dot structure[6,7].

The observed ultralow lasing threshold relies on the high Q cavity mode. This can be further supported by the data taken from the same device with lowered Q-factor, achieved by covering the device with a poly(methyl methacrylate) (PMMA) layer on top. In this situation, the lasing threshold increases up to around 100 μW (Extended Data Fig. 2).

We also study the linewidth evolution around the lasing threshold region. Fig. 2**d** shows the linewidth as a function of output intensity at 160 K. A pronounced "kink" appears around the threshold, similar to the *L-L* curve. Below the threshold, the observed linewidth narrows from ~0.75 nm to ~0.50 nm with increased output power. At the threshold regime, it re-broadens to ~0.65 nm, and then continues to narrow to 0.55 nm. This linewidth dependence is a well-known feature that has been observed in semiconductor nanocavity lasers, such as quantum well[28] and quantum dot nanolasers[7]. The "kink" arises during the phase transition from spontaneous emission into stimulated emission, where the coupling between intensity and phase noise (gain-refractive index coupling) significantly influences the linewidth, and leads to a re-broadened emission spectrum[7,28,29]. Such observed linewidth plateau, together with the *L-L* curve, clearly reveals the lasing behavior in our monolayer semiconductor nanocavity system.

It is essential to lasing that the cavity mode dominates the emission. To illustrate this, we present a contrast experiment between on- and off-cavity regions, by performing a scanning micro-PL measurement on our device. In Fig. 3**a**, we plot a peak distinguishing map, where the normalized peak (739.7nm) height of the lasing spectrum is mapped out over the entire photonic crystal region, indicated by the dashed white line. The dashed orange line indicates the monolayer WSe$_2$, as also shown in the inset SEM device image. The laser emission only comes from the cavity. A set of typical spectra taken on and off cavity (indicated by dashed circles in Fig. 3**a**) is shown in Fig. 3**b**. The on-cavity emission is dominated by the lasing mode while the non-lasing spontaneous emission is prominently suppressed compared to the off-cavity emission. This is strikingly different from the similar observation in the low-Q device where the on-cavity yields similar broad emission from off-cavity[24,25].

Temperature dependent emission behavior of the same device is presented in Fig. 3**c**. A redshift of the emission wavelength with increasing temperature is consistent with the energy shift of the cavity mode[22]. Lasing characteristics are examined at temperatures below 160K, above which the cavity emission is small compared to the background emission and challenging to extract its absolute intensity. Up to about 250 K, the cavity peak diminishes into the recovered background spontaneous emission. A possible explanation of this could be the degradation of the cavity resonance induced by the different thermal expansion between GaP membrane and WSe2 monolayer. Nevertheless, in principle there should be no limitation to increase the lasing temperature. Further improvement of the Q factor by optimizing the cavity designs and fabrication procedures is one way to achieve room temperature lasing. An alternative way is to find other monolayers or their heterostructures that emit photons at energies compatible with silicon photonics. We could then use silicon PhCCs, which have much higher Q than GaP.

We finally discuss the reproducibility of our new lasing architecture based on a monolayer semiconductor and a PhCC. It is routine to fabricate multiple PhCCs on a single chip, while deterministic multiple-transfer of monolayer semiconductors onto different PhCCs can be achieved to make monolayer hybrid devices (Fig. 4**a**). In Fig. 4**b**, we present the lasing spectrum taken from three different devices in the same chip under similar conditions. The lasing devices can be robustly reproduced. It suggests that mass production can be achieved, especially if large area monolayers grown from chemical or physical vapor deposition are utilized.

Our design demonstrates the remarkable possibility to achieve scalable nanolasers using monolayer gain for integrated chip systems. In such a surface geometry, the advantage is that the construction of the optical nanocavity and gain material are naturally separated, allowing fabrication of both parts individually with high quality, before their nondestructive and deterministic combination as hybrids. This unprecedented ability enables their realistic applications in a scalable and designable way, compatible to integrated electronic circuits. Electrically pumped operation and electrostatic tuning the carrier concentration can also be achieved directly, in contrast to the conventional designs. Our monolayer surface-gain geometry presents a versatile lasing technology and an advancement relative to quantum dot nanocavity lasers, with gain material being incorporated after the laser cavity fabrication, which eliminates the degradation of the gain medium during the fabrication process and enables its replacement if needed. We envision that a natural extension of exploring other 2D crystals, as well as their heterostructures, might give a birth to a nanolaser with emission energy compatible to silicon photonics technologies. Many exotic properties of 2D semiconductors may also lead to novel devices such as valley-selective lasers. Beyond nanolasers, many on-chip photonic implementations, such as the study of strongly coupled cavity quantum electrodynamics[30], nonlinear optics and photonic quantum control, could open new horizons based on 2D quantum materials and their heterostructures.

# Figure 1

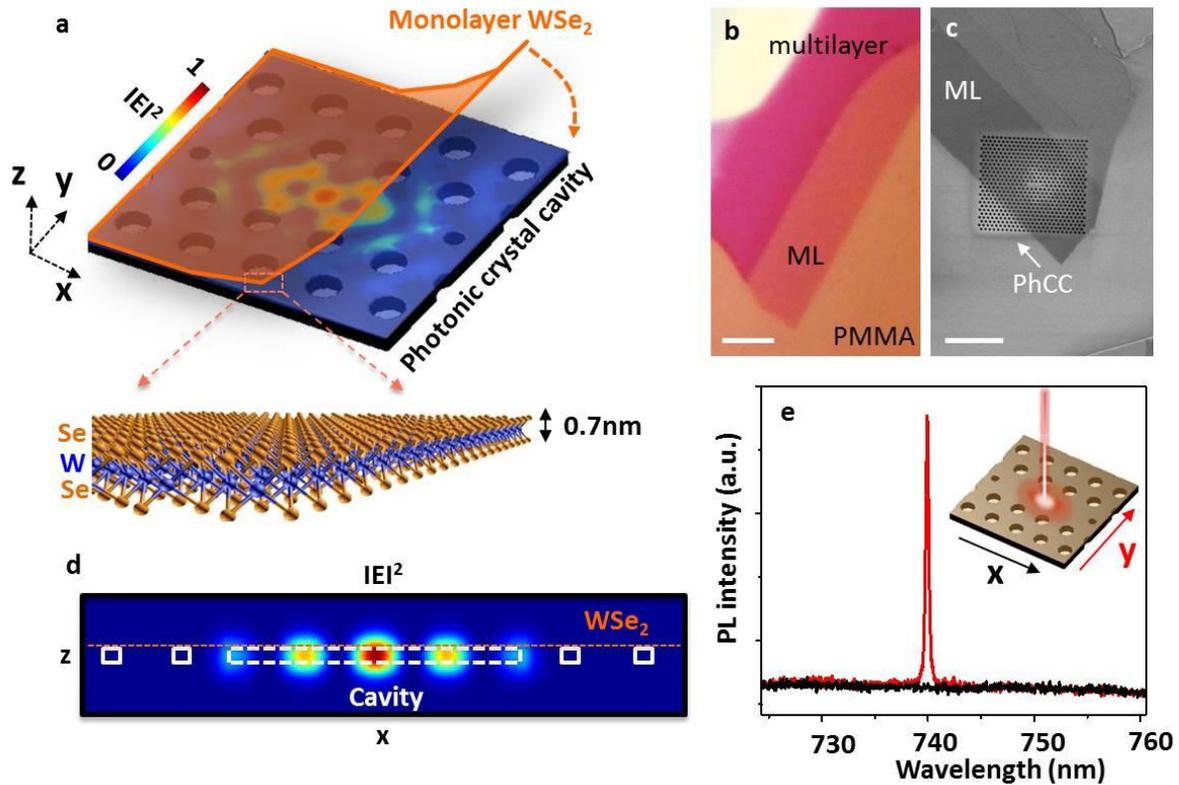

**Figure 1 | Hybrid monolayer WSe$_2$/PhCC nanolasers. a,** Cartoon depiction of our device architecture, where the electric-field profile (in-plane, x-y) of the fundamental cavity mode (pristine cavity before WSe$_2$ transfer) is embedded as the color plot. The inset is a cartoon of the atomic structure of monolayer WSe$_2$. **b,** Optical image of monolayer (labeled as "ML") WSe$_2$ on PMMA before transfer. **c,** SEM image of the hybrid device. Q-factor is ~8000 in this cavity before WSe$_2$ transfer. Scale bars: 3 μm. **d,** Cross section electric-field intensity profile (x-z) of the fundamental mode, where the dashed orange line indicates the ideal position of monolayer WSe$_2$, the solid white rectangles for air holes and dashed white lines for cavity region. **e,** Polarization-resolved PL spectrum of our device taken at 80 K, showing a completely polarized narrow emission at ~740 nm. Black (red) line corresponds to detected linear polarization in x (y) direction.

# Figure 2

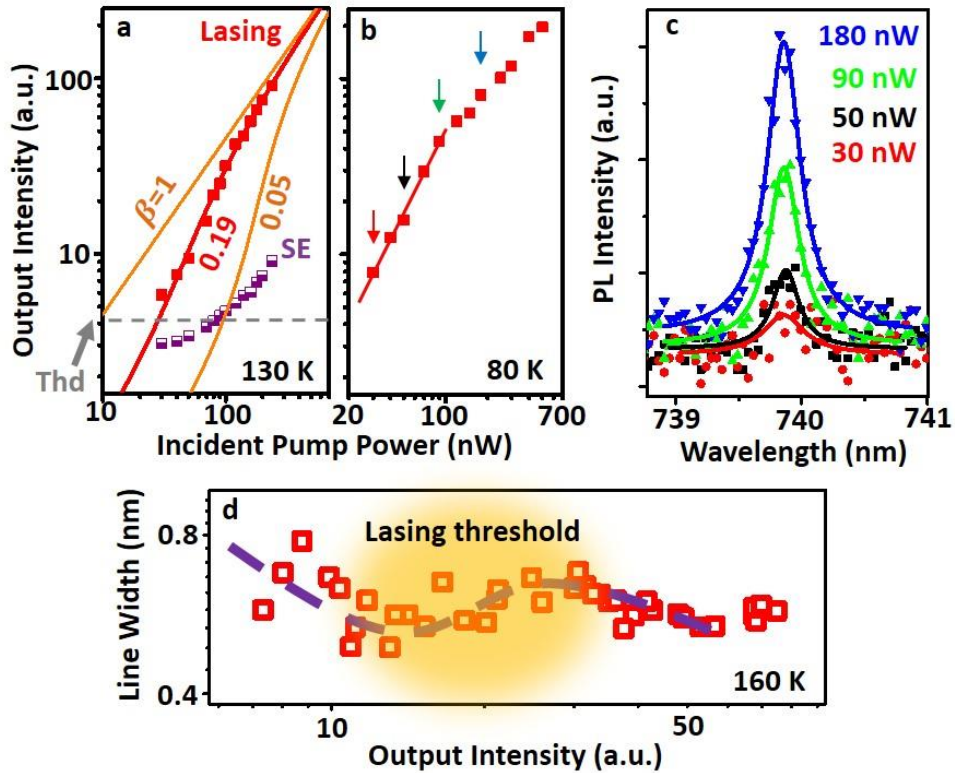

**Figure 2 | Lasing characteristics. a,** Light output intensity (detected power after spectrometer) as a function of the optical pump power (*L-L* curve) at 130 K. Red filled squares correspond to the cavity emission. Violet half-filled square corresponds to the spontaneous emission (SE) off cavity resonance. Solid lines are the simulated curves using the laser rate equation with different $\beta$ factors. $\beta = 0.19$ is the best fit to the lasing data. Dark dashed line corresponds to the defined laser threshold. **b,** *L-L* curve for the same lasing device at 80 K (red squares), where the solid line is a guide for the eye to the transition region. **c,** The PL spectra corresponding to the data points in **b** indicated by the colored arrows. The solid lines are Lorentzian fits to the PL spectra. **d,** Cavity linewidth as a function of the detected output power at 160 K (red empty squares). Dashed line is an eye-guide to the nonlinear linewidth re-broadening area, which corresponds to the lasing threshold region.

# Figure 3

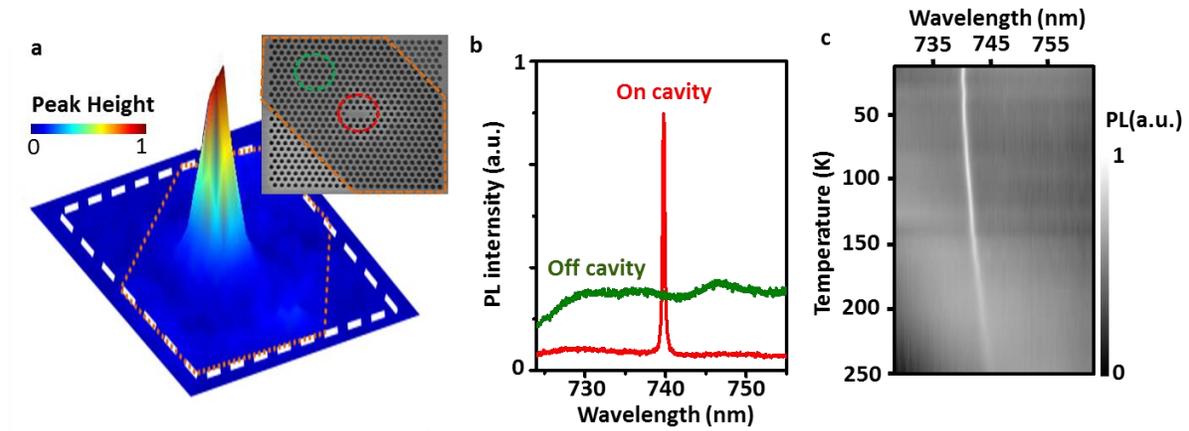

**Figure 3 | Spatially resolved emission and temperature-dependent device behavior. a,** Peak distinguishing spatial map of our device, where peak height, i.e., normalized intensity difference between peak summit (739.7 nm) and bottom (738 nm), is mapped out at 80 K. Dashed white line indicates the photonic crystal area and the dashed orange line shows the area that is covered by monolayer $WSe_2$. Inset is the corresponding device image in SEM. **b,** PL spectra taken on (red) and off (green) cavity region, indicated by the dashed color circles in the inset of **a**. **c,** Temperature dependence of the device emission spectra in grey-scale map.

# Figure 4

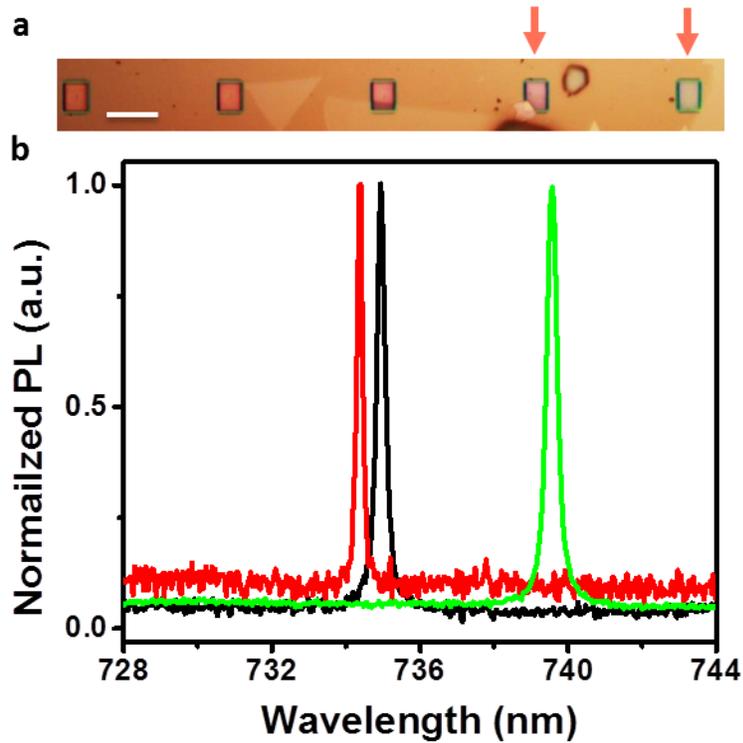

**Figure 4 | Reproducibility and scalability of the 2D nano-lasers. a,** An example of deterministic fabrication of multiple devices on one chip. Here we show an optical image of a typical area with 5 PhCC devices in a row and the last two (indicated by the arrows) are covered with monolayer $WSe_2$. Scale bar: 10 μm. **b,** Lasing spectra (three) can be reproducibly taken from different devices on the same chip under similar conditions.

**Acknowledgments:**

This work was mainly supported by AFOSR (FA9550-14-1-0277). AM is supported by NSF-EFRI-1433496. PhC fabrication was performed in part at the Stanford Nanofabrication Facility of NNIN supported by the National Science Foundation under Grant No. ECS-9731293, and at the Stanford Nano Center. SW was partially supported by the State of Washington through the University of Washington Clean Energy Institute. SB and JV were supported by the Presidential Early Award for Scientists and Engineers (PECASE) administered through the Office of Naval Research, under grant number N00014-08-1-0561. SB was also supported by a Stanford Graduate Fellowship. JY and DGM were supported by US DoE, BES, Materials Sciences and Engineering Division. FH acknowledges the European Commission (FP7-ICT-2013-613024-GRASP).

**Author Contributions**:

XX and AM conceived the experiments. SB and AM fabricated and characterized PhCCs under JV's supervision. SW fabricated the hybrid devices and performed the measurements with assistance from JS and LF, under XX's supervision. SW, XX, AM and SB analyzed the data with discussion from WY and JV. JY and DGM provided the bulk $WSe_2$. FH grew the GaP membrane. SW wrote the paper with input from all authors.

**Author Information**

The authors declare no competing financial interests. Readers are welcome to comment on the online version of the paper. Correspondence and requests for materials should be addressed to X.X. (xuxd@uw.edu ) or A.M. (arka@uw.edu).

# Methods

**Purcell factor estimation**

We estimate the maximum achievable Purcell factor of the cavity, i.e., the peak enhancement of the emission rate, through:

$$F_{max} = \frac{3}{4\pi^2} \frac{Q}{V} \left(\frac{\lambda_c}{n}\right)^3.$$

Here, $F_{max}$ is the maximum Purcell factor, $Q$ is the cavity quality factor, $V \sim \left(\frac{\lambda_c}{n}\right)^3$ is the mode volume, $n \sim 3.1$ is the GaP refractive index and $\lambda_c \sim 740$ nm is the cavity emission wavelength. We obtain $F_{max} \sim 607$ for the as-fabricated cavity $Q = 8000$.

$Q$ factor can be smaller after the monolayer transfer. At room temperature, the Q factor measured after monolayer transfer reduces to $\sim 1300$, consistent with the PL emission at high temperatures. When cooling down to low temperatures, the Q factor recovers to $\sim 2500$. The spatial displacement (z direction), due to the surface-gain geometry, and the random dipole directions of the emitter could also affect the enhancement of the spontaneous emission rate. Considering these effects, the Purcell factor should be written as:

$$F = F_{max} |\psi(s)|^2 <\cos^2\xi>.$$

Here $|\psi(s)|^2 = \left|\frac{E(s)}{E_{max}}\right|^2 \sim 0.4$ is the field intensity ratio (Fig. 1d) between the surface and the central maximum of the cavity, describing the effect of spatial detuning. $\xi$ is the angle between emitter dipole direction (random in x-y plane) and the electric field polarization (y direction). $<\cos^2\xi> = \frac{1}{2\pi}\int_0^{2\pi} \cos^2\xi \, d\xi = \frac{1}{2}$. Therefore we estimate the Purcell factor as F ~ 37 for Q ~ 2500, where we consider the monolayer exciton that is spectrally tuned on the cavity resonance and located right above the center of the cavity.

In real situation, this value could be further reduced. For example, we may also need to consider the spatial displacement of exciton in lateral directions, which will require the knowledge of in-plane exciton distribution that is yet unknown. Moreover, spectral fluctuations of the excitonic linewidth would lead to variation in the Purcell factor over time described by the Lorentzian of the cavity spectrum. However, the spontaneous emission coupling factor $\beta$ is estimated to be ~

0.19 from the measurement (see next section and Fig. 2a of main text), reflecting an efficient Purcell enhancement in this geometry.

**Laser rate equation**

The spontaneous emission coupling factor $\beta$ is an essential figure of merit for a nanocavity laser. To extract its value, we use a rate equation[6] model to describe the evolution of carrier (exciton) number $N$ and the cavity photon number $P$ in the monolayer-PhCC system:

$$\dot{N} = R_{ex} - \frac{N}{\tau_{SE}} - \frac{aNP}{\tau_{cav}},$$

$$\dot{P} = -\frac{P}{t_c} + \Gamma \frac{N}{\tau_{cav}} + \Gamma \frac{aNP}{\tau_{cav}},$$

$$\beta = \frac{\tau_{SE}}{\tau_{cav}}$$

Here, $R_{ex}$ is the optical pumping rate. $\tau_{SE}^{-1}$ is the total spontaneous emission rate. $\tau_{cav}^{-1}$ is the emission rate into the cavity mode. $t_c^{-1}$ is the cavity photon decay rate. $aNP$ is the stimulated emission, which is proportional to $N \cdot P$ with coefficient $a$. $\Gamma$ is the cavity confinement factor. We have ignored non-radiative relaxation processes. The rate of non-radiative decay in monolayer semiconductors is currently not known. Any non-radiative decay would induce additional loss which would result in a larger $\beta$ factor[31]. The transparent carrier number is set to be zero, since it does not affect the fitting result significantly.

We set $\dot{N} = 0$ and $\dot{P} = 0$ to obtain the steady state solution of above coupled equations. The solution is:

$$R_{ex} = \frac{P}{\Gamma t_c (1 + aP)} \left(\frac{1}{\beta} + aP\right)$$

The lasing threshold is defined as the condition when the stimulated emission is equal to the spontaneous emission in the cavity, i.e., $aP = 1$. When $aP > 1$, stimulated emission dominates in the hybrid system and lasing behavior occurs.

We fit our experimental L-L curve with above equation, as plotted in Fig. 2a in the main text. $\beta = 0.19$ is found to be the best fit to the data taken at 130 K.

**PhCC fabrication**:

To fabricate the photonic crystal structures, a 125 nm thick GaP membrane was grown on the top of a 1 μm thick sacrificial $Al_{0.8}Ga_{0.2}P$ layer on a GaP wafer via gas-source molecular beam epitaxy (GSMBE). The patterns were first defined in ZEP520 resist by electron-beam lithography (JEOL JBX 6300, 100 keV) and then transferred to the GaP membrane by a chlorine-based reactive ion etch. Excess resist was removed with Microposit remover 1165 followed by oxygen plasma. The sacrificial layer was finally undercut with hydrofluoric acid to yield suspended membrane structures with high index contrast, followed by cleaning in dilute KOH to remove any by-products of the undercut.

**Hybrid device fabrication**:

The PhCC/$WSe_2$ hybrid structure was fabricated through a standard polymer micro-transfer process. A monolayer $WSe_2$ was first mechanically exfoliated onto a polymer-coated silicon substrate where water-soluble polyvinyl alcohol (PVA, 1%) followed by poly(methyl methacrylate) (PMMA, 950, 6%) was spin-coated on the chip. The stacked monolayer $WSe_2$/PMMA/PVA/Si substrate was then placed on top of water, dissolving the PVA layer to separate the silicon substrate. The floating $WSe_2$/PMMA membrane was transferred using a "perfect loop" (TED PELLA, INC), placing the monolayer onto the pre-fabricated PhCC under microscope followed by heating. The PMMA cover layer was dissolved by a 2-hour acetone bath and a 2-minutes isopropyl alcohol bath.

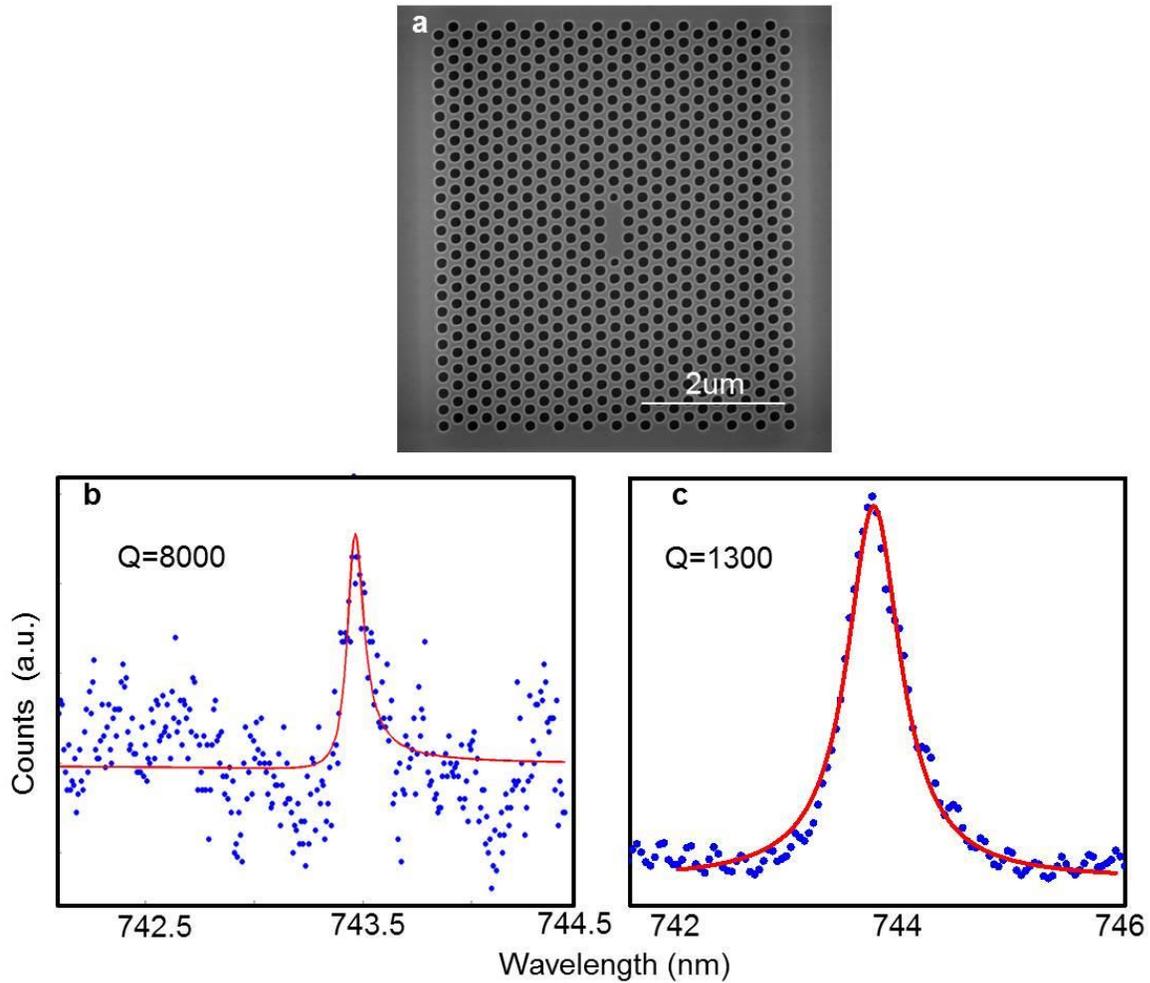

**Extended Data Figure 1 | Cavity Q factor determination. a**, SEM image of a typical PhCC. **b** and **c**, Room temperature cross-polarized reflection taken from this cavity, before (**b**) and after (**c**) monolayer WSe$_2$ transfer. The as-fabricated cavity (before transfer) of our lasing devices typically have Q-factors ranging from 5000 to 14000. After monolayer transfer, the Q factor is reduced from 8000 to 1300 in this device. After cooling down to cryogenic temperatures, the Q factor recovers to ~ 2500.

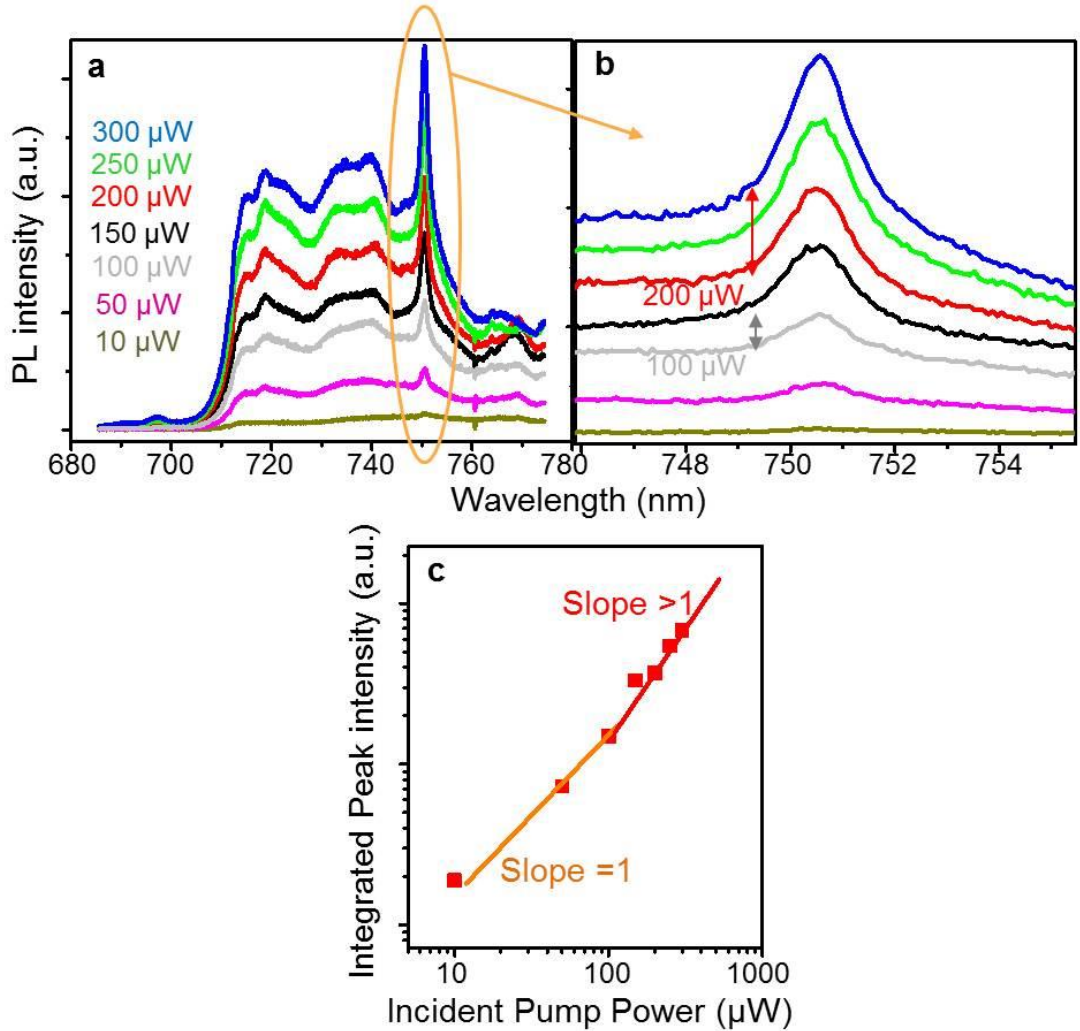

**Extended Data Figure 2 | Device behavior with poly(methyl methacrylate) (PMMA) reduced Q-factor. a**, PL spectra taken from the PMMA covered device at different pumping powers (30 K), showing pronounced cavity peaks. **b**, A zoom in of the cavity peaks. **c**, Power dependence of the integrated peak-intensity. A nonlinear "kink" appears around 100 μW. The PMMA layer reduces Q-factor to ~ 500, and also shifts the resonance to lower energy (750.7nm). This supports the conclusion that the ultralow lasing threshold in our device results from the high Q-factor, by significantly enhancing the spontaneous emission rate into the lasing mode.

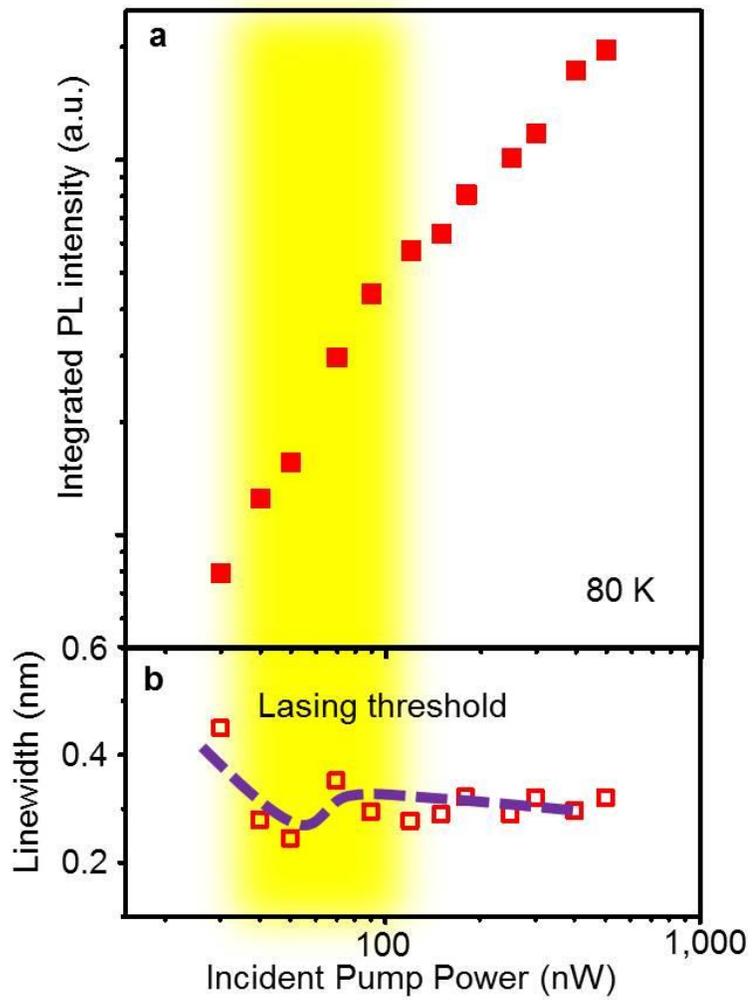

**Extended Data Figure 3 | The nonlinear "kinks" at 80 K**. Both integrated emission intensity (**a**) and linewidth (**b**) are shown. The same set of data as in Fig. 2**b**.